\documentclass[aps,twocolumn,amsfonts,showpacs]{revtex4}

\begin{document}
\newcommand{\bb}{\begin{equation}}
\newcommand{\ee}{\end{equation}}
\newcommand{\eqb}{\begin{eqnarray}}
\newcommand{\eqf}{\end{eqnarray}}

\preprint{DRAFT}
\title{Gauge Field Theory in the Infrared Regime }
\author{Ashok Das}
\email{das@pas.rochester.edu}
\affiliation{Departament of Physics, University of Rochester,
Rochester, N.Y. USA}
\author{J. Gamboa}
\email{jgamboa@lauca.usach.cl}
\affiliation{Departamento de Fisica, Universidad de Santiago de Chile,
Casilla 307, Santiago, Chile}
\author{J. L\'opez-Sarri\'on}
\email{justo@dftuz.unizar.es}
\affiliation{Departamento de F\'{\i}sica, Universidad de Santiago
de Chile, Casilla 307, Santiago, Chile}
\author{F. A. Schaposnik}
\email{fidel@fisica.unlp.edu.ar}\thanks{Affiliated to CICBA, Argentina}
\affiliation{Departamento de F\'\i sica, Facultad de
Ciencias Exactas \\Universidad Nacional de La Plata, CICBA and IFLP,
 Argentina}
\begin{abstract}
We propose that the low energy behavior of a pure gauge theory can be
studied by simply assuming violation of Lorentz invariance which is
implemented through a deformation of the canonical Poisson brackets of
the theory depending on an infrared scale. The resulting theory is
equivalent to a pure gauge theory with a Chern-Simons like term. It
is shown that at low energies this theory can be identified with three
dimensional QCD where the mass of the fermion is related to
the infrared scale.
\end{abstract}
 \maketitle

The framework of relativistic quantum field theory has been
extraordinarily successful in explaining various quantum mechanical
phenomena over a wide range of energies. The agreement of the
experimental measurements with the results of perturbative calculation
for $g-2$ has been remarkable. The predictions of the standard model
have also been tested to a very high degree of accuracy and one
believes that the discovery of the Higgs particle may be just around
the corner.

In spite of all the success, one does not believe that relativistic
quantum field theory, in its present form, will be applicable at all
energies. There is, of course, the suggestion that at very high
energies, string theory will describe all the fundamental interactions
of nature in which case we can think of relativistic quantum field
theory only as a low energy (compared to the Planck scale) effective
field theory. Even if, one does not believe in string theory, it is
recognized that in order to be compatible with several gravitational
effects and astrophysical observations, relativistic quantum field
theory must be modified to incorporate a tiny violation of Lorentz
invariance which has been the cornerstone of much of twentieth century
physics \cite{vuce,koste}. At low energies, clearly we do not expect a
Lorentz invariant 
description either. Therefore, there must exist an infrared scale
below which a nonrelativistic description should be more
meaningful. For example, in an Abelian gauge theory such as QED, at
energies much smaller than the electron mass, the theory of electrons
interacting with photons can be conveniently described by an effective
field theory which leads to a nonrelativistic
Schr\"{o}dinger equation (the photon is always relativistic). Such a
procedure, however, cannot be readily carried out in the case of a
non-Abelian gauge theory such as QCD which below a certain energy scale
(roughly the proton mass) becomes confining. Namely, in this case,
in the infrared limit, a nontrivial physical picture arises
which necessitates the use of  nonperturbative methods since  phenomena
such as confinement and hadronization are beyond perturbation
theory. Unfortunately, nonperturbative techniques are not very well
understood as yet.

Precisely in this regime, however, there is a remarkable analogy
between molecular physics and effective heavy quark theory following
from QCD. Indeed, in molecular physics it is reasonable to assume (as
a first approximation) that  nuclei are much heavier compared to the
electrons and, therefore, are static and as a  first approximation,
are decoupled from electrons  \cite{wilc}. On the other hand, this
assumption (known as the Born-Oppenheimer approximation) cannot be
completely right since, as is well known,
the electronic and the nuclear variables are coupled through Berry's phase
 \cite{berry}. Following  Berry's analysis \cite{berry,wilc},
one finds that at low energies a gauge symmetry emerges
as a consequence of global geometrical considerations.
This symmetry induces new dynamical effects on the spectrum
which has been verified  in numerous experiments \cite{exp}.

Along the same lines one could think that  the heavy quark
effective field theory \cite{heavy} (where the heavy quarks are
simply assumed to be spectators) should necessarily contain
corrections due to the interactions with light quarks as in the
case of molecular physics \cite{wilc}. Although it is technically
difficult at the present time to go beyond the Born-Oppenheimer
approximation in the context of the effective heavy quark field
theory, it seems reasonable to explore other possibilities simply
by assuming that, in this regime, Lorentz invariance is expected to be
violated
as a consequence of the infrared scale. Since light quarks are not
being integrated out in this approach, the situation may appear to
be different from that in molecular physics, nonetheless the
approach may represent a step forward in the context of effective
heavy quark field theory.

One way to introduce violation of Lorentz invariance into a field
theory is by deforming  the canonical commutation relations of the
theory (so as to have a noncommutative field theory). In two recent papers
\cite{justo,horacio} such a noncommutative gauge field theory
violating Lorentz invariance has been studied. The idea is quite
simple. One starts with the standard action for a gauge field
theory given by
\bb
S = -\frac{1}{4} \int d^4x\ F_{\mu\nu}^{a} F^{\mu\nu a},\label{h11}
\ee
where the field strength tensor is defined by
\begin{equation}
F_{\mu\nu}^{a} = \partial_{\mu}A_{\nu}^{a} - \partial_{\nu}A_{\mu}^{a}
+ g f^{abc} A_{\mu}^{b} A_{\nu}^{c},
\end{equation}
with $f^{abc}$ denoting the structure constants of the group. In the
context of a noncommutative field approach (inspired by noncommutative
geometry), one modifies the conventional canonical (equal-time)
Poisson bracket relations on the phase space as
\begin{widetext}
\begin{eqnarray}
\{A^a_{\mu}(x),A_{\nu}^b(x') \} &=& 0\rightarrow
\{A^a_{\mu}(x),A_{\nu}^b(x') \} = 0 , \nonumber\\
\{A^a_{\mu}(x),\Pi_{\nu}^b(x') \} &=&\delta^{ab}
\eta_{\mu\nu}\delta^3(x-x')\rightarrow \{A^a_{\mu}(x),\Pi_{\nu}^b(x')
\} = \delta^{ab} \eta_{\mu\nu}\delta^3(x-x'),  \nonumber\\
\{\Pi_{0}^{a} (x), \Pi_{\mu}^{b} (x')\} & = & 0\rightarrow
\{\Pi_{0}^{a} (x), \Pi_{\mu}^{b} (x')\} = 0,\nonumber\\
\{\Pi^a_i(x),\Pi_j^b(x') \} &=& 0\rightarrow \{\Pi^a_i(x),\Pi_j^b(x')
\} = \epsilon_{ijk}\theta_k\delta^{ab} \delta^3(x-x'),
\label{modified}
\end{eqnarray}
\end{widetext}
where we have introduced a constant space-like deformation parameter
$\theta^{\mu} = (0, \vec{\theta}\ )$ which has the canonical dimension
 1. This modification of the canonical brackets clearly  breaks
Lorentz invariance and since we would expect a violation of Lorentz
invariance only in the infrared sector of the theory, we would like to
identify the magnitude of this vector as defining the infrared
scale.

It should be noted that the modification of the Poisson bracket
relations preserves gauge invariance. Indeed,   it can be seen
with a little bit of algebra that the theory (\ref{h11}) with the
modified Poisson brackets (\ref{modified}) is equivalent to the
theory (with the naive canonical Poisson brackets) described by
the action \bb S=\int d^4x \left[ -\frac{1}{4}
F_{\mu\nu}^{a}F^{\mu\nu a}
  +\theta^\mu C_\mu\right], \label{int1}
\ee
where $\theta^\mu$ is the constant four-vector introduced earlier (as
the deformation parameter) and $C_\mu$ is given by
\begin{equation}
C_\mu = \epsilon_{\mu \nu \lambda \rho} \left[A^{\nu a}
\partial^\lambda A^{\rho a} + \frac{2g}{3} f^{abc} A^{\nu a}
A^{\lambda b} A^{\rho c} \right].
\end{equation}
From the form of this action it is clear that while it violates
Lorentz invariance (because of the specific form of
$\theta^{\mu}$), it is manifestly gauge invariant. However, because of
the additional term, the
conventional Gauss' law ($D_{i}$ denotes the covariant derivative)
\bb
(D_{i} E_i)^{a} = (D_{i}F_{0i})^{a} = 0, \label{gauss1}
\ee
 is
replaced by in this case by
 \bb
(D_{i}F_{0i})^{a} + \epsilon_{ijk} \theta_{i} F_{jk}^{a} =  (D_{i}
 E_i)^{a} + \theta_i B_{i}^a =0.
\label{gauss2}
 \ee
 Furthermore, it is clear that if we choose
$\theta^{\mu} = (0,0,0,\theta)$, then the $\theta^\mu C_{\mu}$ term has
the form of the three dimensional Chern-Simons term \cite{DJ}
(although the fields live in four dimensions). It should be
stressed that the approach discussed above is consistent with that
proposed in \cite{carroll}, in which Lorentz invariance is
essentially broken through the addition of the Chern-Simons term
to a Maxwell or  Yang-Mills theory coupled to an external four
vector field.

The purpose of this short note is to propose  the model defined
by the action (\ref{int1}) as an example of QCD at low energy (low
compared to the scale $\theta$) within an alternative approach. We
note that one can try to derive (\ref{int1})  as an effective
theory by introducing fermion fields such that when they are
integrated out one ends with a Lorentz and
 CPT violating Chern-Simons
term. This has been achieved in \cite{Colladay}-\cite{Chung} by
including  a coupling to an axial-vector term which violates
Lorentz invariance in an extended Dirac Lagrangian. In fact,
originally, the interest in theories containing a Chern-Simons
term was triggered by the observation that such a term is induced in
the effective action of the gauge field through fluctuations of
massive fermion fields. This is the celebrated parity anomaly
 \cite{DJ, red}.
Another possibility was discussed in \cite{RedlichW} by studying a
theory  with $SU(2)_L$ gauge fields and a finite chemical
potential term which can be interpreted as the coupling of an
external $U(1)$ gauge field to a $U(1)$ current.  Then,  as a
consequence of the $U(1)_L$ anomaly of such a current, a
Chern-Simons term is generated. In the following, we will take an
approach related to this latter case.

Let as add  to the Lagrangian density of QCD a gauge invariant chiral
fermion term of the form
\[
{\cal L}_{\mbox{f}} = i {\bar \psi}_L  {D \hspace{-.6em} \slash
\hspace{.15em}} \psi_L \, ,
\]
with $\psi_L$  a left handed Weyl spinor, so that the complete
Lagrangian density reads
\bb {\cal L} = -\frac{1}{4} F_{\mu\nu}^a F^{\mu\nu a} + i {\bar
\psi}_L {D \hspace{-.6em} \slash \hspace{.15em}} \psi_L.
\label{weyl} \ee
Integrating out  the chiral fermions would yield an effective action
of the form
\bb
{\cal L} = - \frac{1}{4} F_{\mu\nu}^{a} F^{\mu\nu a} + \log \det
 \left[ i {D \hspace{-.6em} \slash \hspace{.15em}} \right].
 \label{weyl1}
\ee
Now, for Weyl fermions the Dirac operator does not have a well
defined eigenvalue problem since it maps negative chirality
spinors into positive chirality ones. This implies that the
definition of the fermion determinant in (\ref{weyl1}) is, in
general, problematic. One way out is to modify the gauge covariant Dirac
operator by adding free right-handed fermions into the theory so
that one ends up with
 \bb
D\!\!\!\!\slash = \left(\partial\!\!\!\slash - ig
A\!\!\!\slash\right) P_{-}\rightarrow   D[A] =  \partial\!\!\!\slash
- ig A\!\!\!\slash P_- ,
\label{modified1}
\ee
where $P_{-}$ denotes the projection operator onto the left-handed spinor
space
\begin{equation}
P_- = \frac{1-\gamma_5}{2}.
\end{equation}
The operator $ D[A]$ does define an
eigenvalue problem since it acts on Dirac fermions {\it i.e.}
\bb i
D[A] \varphi_n = \lambda_n \varphi_n, \label{eig1}
\ee
 so that the
fermion determinant can be computed as the product of eigenvalues
in the usual way. However, since the Dirac operator is unbounded, the
product of eigenvalues
needs to be regularized. Furthermore, since the right-handed
fermions are decoupled from the
gauge field, the modified operator
(\ref{modified1}) is not gauge covariant as the original one was.
Hence, there is no way to obtain a gauge invariant answer
for the regularized fermion determinant. It is
through this mechanism that an anomalous term is generated
(see for example \cite{anomal}
and references therein).

Let us note that (we use the convention
$\left(\gamma^{0}\right)^{\dagger} = \gamma^{0},
\left(\gamma^{i}\right)^{\dagger}=-\gamma^{i},\gamma_{5}=i\gamma^{0}
\gamma^{1}\gamma^{2}\gamma^{3} = \left(\gamma_{5}\right)^{\dagger}$) 
\begin{equation}
\left(\gamma_{5}\gamma^{3}\right)^{\dagger} =
\gamma_{5}\gamma^{3},\quad \left(\gamma_{5}\gamma^{3}\right)^{2} =
1,\quad \det \left(\gamma_{5}\gamma^{3}\right)=1,
\end{equation}
so that we can write
\begin{equation}
\det \left(i D[A]\right) = \det \left(i\gamma_{5}\gamma^{3}
D[A]\right).
\end{equation}

Returning to our discussion of the infrared sector of QCD let us
choose, as suggested earlier, $\theta^\mu = (0,0,0,\theta)$ in
(\ref{int1}). Moreover,
we partially  fix the gauge degrees of freedom through the
condition $A_3 =0$. With this,
the eigenfunctions (\ref{eig1}) can be expanded in an interval $0\leq
x^{3}\leq 1/\theta$ with anti-periodic boundary
condition\footnote{Periodic boundary conditions for fermions are known
  to lead to acausal propagation for gauge fields which is why we use
  anti-periodic boundary condition. We refer the reader to L. H. Ford,
  Phys. Rev. {\bf D20}, 933 (1980), A. Das and M. Hott,
  Mod. Phys. Lett. {\bf A10}, 893 (1995) for details.} as
\bb
\varphi_n (x^0,x^1,x^2,x^3) =
 \sum_{m=-\infty}^\infty e^{i (2m+1)\pi \theta x^3}
 {\tilde \varphi}_{n,m }(x^0,x^1,x^2). \label{eg2}
\ee
We note that, for a fixed value of $m$, the operator
$i\gamma_{5}\gamma^{3}D[A]$,  in the space of
functions $\tilde{\varphi}_{n,m}$, has the form
\begin{equation}
i\gamma_{5}\gamma^{3} D[A]  = \left(i\gamma_{5}\gamma^{3}\gamma^{\alpha}
(\partial_{\alpha} - ig A_{\alpha} P_-)
+ M(m)\gamma_{5} \right),\label{DA}
\end{equation}
where $\alpha=0,1,2$ and we have identified $M(m)=
(2m+1)\pi\theta$. For simplicity and clarity, let us choose the
Weyl basis (chiral basis) for the representation of the gamma functions,
\begin{equation}
\gamma^{\mu} = \left(\begin{array}{cc}
0 & \sigma_{\mu}\\
\bar{\sigma}_{\mu} & 0
\end{array}\right),\quad \gamma_{5} = \left(\begin{array}{rr}
-I & 0\\
0 & I
\end{array}\right), 
\end{equation}
where $\sigma_{\mu} = (I,\sigma_{i}), \bar{\sigma}_{\mu} = (I,
-\sigma_{i})$ with $I$ denoting the two dimensional unit matrix.
In this representation, Eq. (\ref{DA}) has the explicit form
\begin{eqnarray}
& & i\gamma_{5}\gamma^{3} D[A]\nonumber\\
& = & \left(\begin{array}{cc}
-i\sigma_{3}\bar{\sigma}_{\alpha}\left(\partial_{\alpha}-igA_{\alpha}\right)
- M(m) & 0\\
0 & -i\sigma_{3}\sigma_{\alpha}\partial_{\alpha} + M(m)
\end{array}\right).\nonumber\\
& &
\end{eqnarray}
This shows that the right handed fermions decouple and lead only to a
constant multiplicative factor in the evaluation of the
determinant. Furthermore, identifying the $2\times 2$ Dirac matrices
of three dimensional space-time as
\begin{equation}
\gamma_{(3)}^{\alpha} = -\sigma_{3}\bar{\sigma}_{\alpha},\quad \alpha=0,1,2,
\end{equation}
we obtain, for any fixed $m$,
\begin{eqnarray}
\det \left(iD[A]\right) & = & \det \left(i\gamma_{5}\gamma^{3}
D[A]\right)\nonumber\\ 
& = & {\cal N} \det \left(iD\!\!\!\!\slash^{(3)} - M(m)\right).
\end{eqnarray}

Thus, for energies much lower than the infrared scale, $E\ll \theta$,
only the lowest mode $(m=0)$ contributes and one finds
that the four and the three dimensional determinants coincide up to a
multiplicative factor arising from the decoupled
right handed sector,
\bb
\det \left( i {D \hspace{-.6em} \slash \hspace{.15em}}\right)^{(4)} =
{\cal N}
\det \left(  {i D \hspace{-.6em} \slash \hspace{.15em}} -
\pi\theta \right)^{(3)}, \label{equa1}
\ee
which is an important result for our work. Furthermore, the three
dimensional determinant for a massive fermion (with mass $\pi\theta$) is
known to generate a Chern-Simons term which is how the action
(\ref{int1}) can be obtained as an effective action.

Indeed, at low energies, we can generalize this equivalence by scaling
\eqb
&& \psi_L \rightarrow  \sqrt{\theta} \psi,   \nonumber
\\
&& g^{(4)}  \rightarrow  \left(\sqrt{\theta}\right)^{-1} g^{(3)},
\eqf
 One can then describe the pure gauge theory with a term violating Lorentz
 invariance as in (\ref{int1}), in terms of relativistic three
 dimensional QCD with fermions having a mass related to the infrared
 scale, namely
\bb
{\cal L} = -\frac{1}{4} F_{\mu\nu}^{a} F^{\mu\nu a} + {\bar \psi} \left(i
{D \hspace{-.7em} \slash \hspace{.15em}}^{(3)} - \pi\theta \right)\psi.
\label{qed1}
\ee

In summary, in this paper we have argued that a four dimensional
gauge theory with a Chern-Simons term could be considered as
bosonized QCD at low energy. Although in four dimensions this
theory violates Lorentz invariance, nonetheless one can map the
gauge theory with a Chern-Simons term to relativistic QCD in three
dimensions.  Thus, the point of view advocated in this paper would
suggest a new route for an old problem in the nonperturbative
sector of the gauge theory. We expect to compare the abundant
results in three dimensional QCD \cite{qcd3} with heavy quark
theory \cite{shifman} in a future publication.

This work was supported in part by US DOE Grant number DE-FG
02-91ER40685, FONDECYT 1050114, MECESUP-0108, PIP-CONICET 6160,
and CICBA and UNLP grants.

\end{document}